\documentclass[a4paper]{article}
\usepackage[T1]{fontenc}
\usepackage{lmodern} 

\usepackage{geometry}
\usepackage{authblk}
\usepackage{setspace}
\usepackage{graphicx}
\graphicspath{{./figures/}}
\usepackage{subcaption}
\usepackage{amsmath, amssymb, mathrsfs}
\usepackage{float}
\usepackage{balance}
\usepackage{enumitem, array}
\usepackage{multirow, booktabs}
\usepackage{tabularx}
\usepackage{comment}

\usepackage[colorlinks=true, citecolor=blue, linkcolor=blue, urlcolor=blue]{hyperref}

\usepackage[
    style=numeric-comp,
    citestyle=numeric,
    sorting=none,
    backend=biber
]{biblatex}
\usepackage[labelfont=bf]{caption}

\title{The Norwegian-Polish CCS Network: A Case Study in Bilateral Collaboration for European Climate Action}

\author[1*]{Mohammad Nooraiepour}
\author[2]{Pawel Gladysz}
\author[3]{Eirik Melaaen}

\affil[1]{\small {Department of Geosciences, University of Oslo, P.O. Box 1047 Blindern, 0316, Oslo, Norway.}}
\affil[2]{\small {Faculty of Energy and Fuels, AGH University of Krakow, Mickiewicza 30 Av., 30-059 Krakow, Poland.}}
\affil[2]{\small {Norwegian Energy Partners (NORWEP), Hoffsveien 23, 0275 Oslo, Norway.}}
\affil[*]{\small{Corresponding author: mohammad.nooraiepour@geo.uio.no}}

\begin{document}
\maketitle

\begin{abstract}

In the face of escalating climate change, achieving significant reductions in greenhouse gas emissions from hard-to-abate industrial sectors is imperative. Carbon Capture and Storage (CCS) represents an essential technological advancement to achieve sustainable decarbonization. This manuscript reports on the bilateral CCS network between Norway and Poland, designed and implemented to leverage their capabilities to expedite technology deployment via mutual cooperation and accelerate CCS initiatives targeted to member states' challenges across Europe, aiming for a meaningful contribution to climate goals. Norway is renowned for its operational acumen demonstrated through landmark projects like Sleipner and Snøhvit, and its forward-looking initiatives, such as the open-source cross-border Northern Lights project, offer advanced infrastructure and expertise. Conversely, Poland, characterized by its coal-dependent economy and the challenge of decarbonizing extensive industrial emissions, presents a significant geological CO$_2$ storage potential, estimated at over 15.5 gigatonnes. This study delves into the potential synergies derived from collaborative endeavors in academic education, research and development, industrial implementation, regulatory coherence, and public engagement. By underscoring the reciprocal benefits of such partnership, the study underscores the indispensable role of bilateral cooperation in harnessing CCS's capabilities to meet the EU's ambitious climate objectives, paving the way toward a sustainable and low-carbon future. Additionally, it outlines a scalable model for fostering and supporting broader bilateral and multi-lateral collaborations, emphasizing the pivotal role of interconnected networks in shaping effective global climate action strategies.
\end{abstract}

\section*{Introduction}

The intensifying impacts of climate change, driven by anthropogenic greenhouse gas emissions, necessitate urgent and decisive global action. The Intergovernmental Panel on Climate Change (IPCC) Sixth Assessment Report emphatically emphasizes the need for immediate and substantial emissions reductions across all sectors to meet the target of limiting global warming to well below 2°C, with aspirations to confine it to 1.5°C above pre-industrial levels \cite{kikstra2022ipcc}. Carbon dioxide (CO$_2$), as the most prevalent long-lived greenhouse gas, is a primary driver of this challenge, necessitating targeted mitigation strategies \cite{ussiri2017carbon}. Within this context, Carbon Capture and Storage (CCS) emerges as a critical technological intervention for mitigating CO$_2$ emissions, particularly from energy-intensive and hard-to-abate industrial sectors \cite{paltsev2021hard, kikstra2022ipcc}. These sectors include power generation (notably coal and gas-fired plants), cement production (approximately 8\% of global CO$_2$ emissions), steel manufacturing (7-9\% of global emissions), as well as refineries and chemical production facilities \cite{paltsev2021hard, kumar2024decarbonizing}.

CCS encompasses a suite of complementary technologies that capture CO$_2$ emissions at their source, preventing their release into the atmosphere. Once captured, the CO$_2$ is purified, compressed, and transported via pipelines, ships, or other means to suitable geological formations for permanent storage. These formations typically include primarily deep saline aquifers and then depleted oil and gas reservoirs suitable for long-term containment and where CO$_2$ can effectively be sequestered \cite{ringrose2021storage}. CCS offers an actionable pathway for decarbonizing existing industrial facilities to extend their operational lifespan while significantly reducing their emissions and environmental footprint. Furthermore, CCS is indispensable in producing low-carbon hydrogen (blue hydrogen) through steam methane reforming of natural gas, whereby the generated CO$_2$ is captured and stored \cite{massarweh2023blue, bauer2022climate}. This low-carbon hydrogen serves a myriad of applications, including transportation, heating, and industrial processes, underpinning a broader transition to a sustainable energy economy.

The European Union (EU) has articulated ambitious climate targets, aiming for climate neutrality by 2050 as part of the European Green Deal \cite{wolf2021european}. Realizing this visionary goal necessitates a profound transformation of the energy system and the widespread adoption of low-carbon technologies, with the capture-sequestration component poised to play a pivotal role \cite{budinis2018assessment}. Recognizing this, the EU has integrated CCS into its climate and energy policies, including instruments such as the EU Emissions Trading System (ETS), which offers a market-based mechanism for carbon pricing, theoretically incentivizing CCS adoption \cite{teixido2019impact, rickels2020future}. Additionally, the EU Innovation Fund supports deploying cutting-edge low-carbon technologies, with CCS as a focal area \cite{mikunda2020briefing, rienks2023eu}. Despite this strategic positioning, the large-scale deployment of CCS within Europe remains burdened by challenges, including high capital costs associated with capture technologies, transport infrastructures, public acceptance concerns over perceived risks and long-term storage safety, as well as the need for robust, harmonized regulatory frameworks across member states \cite{budinis2018assessment, weber2018uncertain}.

Norway, distinguished as a global leader in CCS, has demonstrated the technical viability and long-term safety of geological CO$_2$ storage through pioneering projects \cite{ringrose2020store, furre2019building}. The Sleipner project, operated by Equinor since 1996, has successfully stored over 20 million tonnes of CO$_2$ in the Utsira Formation, a deep saline aquifer beneath the North Sea \cite{ringrose2021storage, furre2019building}. This project not only demonstrated the large-scale feasibility of geological storage but also yielded invaluable scientific data on CO$_2$ migration and containment \cite{furre2024sleipner}. Similarly, the Snøhvit project, operational since 2008, captures CO$_2$ from natural gas processing at the Melkøya LNG plant, subsequently injecting it into a subsea reservoir near the Snøhvit gas field in the Barents Sea, sequestering approximately 0.7 million tonnes of CO$_2$ annually \cite{hansen2013snohvit, ringrose2021storage}. These initiatives have provided significant operational experience, scientific insights, and contributed to establishing public confidence. The Northern Lights/LongShip project further advances the CCS frontier by developing a pioneering open-access CO$_2$ transport and storage infrastructure designed to receive CO$_2$ from industrial emitters across Europe \cite{furre2019building, meneguolo2024subsurface}. With an expected initial injection capacity of 1.5 million tonnes annually, and potential expansion to over 5 million tonnes, Northern Lights symbolizes the scaling potential of CCS in Europe. Norway’s comprehensive regulatory framework, aligned with the EU CCS Directive and supplemented by national legislation, provides a stable and predictable investment landscape for CCS projects \cite{dixon2015legal,romasheva2019ccs}.

In contrast, Poland faces the complex challenge of transitioning from a heavily coal-dependent energy infrastructure. Coal currently generates a significant fraction of Poland's electricity (approximately 60\% in 2023 as shown in Fig. \ref{fig:figure1}), underpinning various industrial processes and resulting in substantial CO$_2$ emissions \cite{pluta2023scenario, gladysz2020techno}. The Bełchatów Power Station, Europe’s largest lignite-fired power plant with a capacity exceeding 5 GW, epitomizes the scale of industrial emissions in Poland \cite{zak2022energy}. Transitioning Poland's energy sector to align with both national and EU climate commitments is therefore essential. Nevertheless, Poland possesses robust geological CO$_2$ storage potential, estimated at over 15 Gt \cite{CLAT2025}, concentrated in deep saline aquifers within the Polish Lowlands and depleted hydrocarbon reservoirs in the Carpathian Foredeep \cite{koteras2020assessment, rutters2021state}. This extensive storage capacity, coupled with urgent reduction needs for industrial emissions, positions Poland as a critical partner in a broader European CCS network. 

Poland's CCS development is relatively nascent compared to Norway's. Therefore, Poland requires significant investments in infrastructure, technology adaptation, and regulatory framework establishment. To expedite deployment and effectively implement strategies for success, Poland needs to partner with a competent player with extensive experience in diverse aspects of CCS. Through strategic partnerships, Poland can leverage existing expertise, reduce initial barriers to implementation, and position itself as a proactive player in the broader European CCS network, ultimately contributing to national and regional decarbonization goals.

A bilateral collaboration between Norway and Poland presents a strategic opportunity to leverage their complementary strengths to expedite CCS deployment in both nations and across Europe. Norway's expertise, operational know-how, advanced infrastructure—including the open-access Northern Lights project—and established regulatory framework could significantly mitigate barriers to CCS implementation in Poland, thereby reducing initial investments and operational risks. Conversely, Poland's significant storage potential and national resources offer a long-term solution for CO$_2$ sequestration in the region. Such a partnership may also facilitate valuable technology transfer, foster innovation through knowledge exchange, enhance technical capabilities, enable joint R\&D and industrial initiatives, align regulatory requirements, and establish shared infrastructure, collectively reducing costs through economies of scale and accelerating the European CCS implementation.

This manuscript provides a comprehensive overview of the current state of CCS deployment and investigates the synergistic opportunities presented by establishing a Norwegian-Polish CCS network within these two countries. It critically evaluates the benefits of collaboration across multiple dimensions, including research and development, technological advancement, industrial implementation, regulatory harmonization, and public engagement. By illustrating the potential of this bilateral partnership, the study emphasizes its vital role in advancing CCS deployment, thereby contributing to a sustainable, low-carbon future. Additionally, this bilateral model exhibits scalability, indicating its potential to support broader collaborations. It advocates for the development of interconnected, multilateral CCS networks, which could serve as a pivotal strategy to enhance global climate action effectiveness and facilitate comprehensive value chain development.

\section{Poland at the Decarbonization Crossroad}

\begin{figure}
    \centering
    \includegraphics[width=0.9\textwidth]{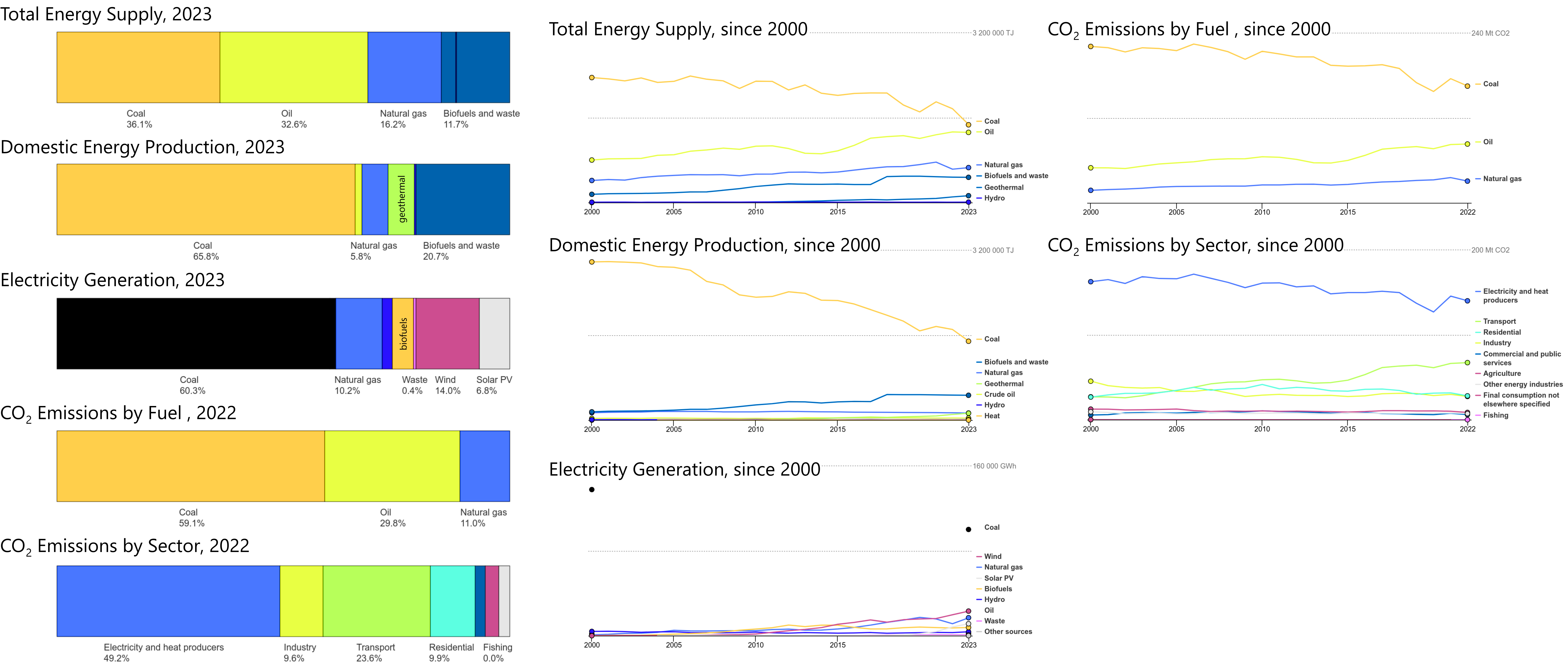}
    \caption{Overview of Poland's energy mix and CO$_2$ emissions for the years 2023/2022 and historical trends since 2000. The figure comprises several subplots: 
    (a \& f) Total Energy Supply: This includes all energy produced domestically or imported, minus exports and storage. It represents the total energy available to meet domestic end-user demands. 
    (b \& g) Domestic Energy Production: Encompasses fossil fuels that can be burned for electricity or used as fuels, alongside renewable energy sources like hydro, wind, and solar photovoltaic (PV). Bioenergy, comprising both modern and traditional sources, including the incineration of municipal waste, is also considered. 
    (c \& h) Electricity Generation: Highlights the transformation into electricity, primarily through thermal power plants—where energy losses occur—and renewable sources that harness natural forces like the sun, wind, or hydropower. 
    (d \& i) CO$_2$ Emissions by Fuel: Details emissions arising from fossil fuel combustion, such as coal, oil, and natural gas, used for power generation and fueling vehicles and machinery. 
    (e \& j) CO$_2$ Emissions by Sector: Provides a sectoral breakdown of energy-related emissions, influenced by economic and energy system structures. This includes emissions from burning fuels in power plants, oil-based fuels in transportation, fossil fuel heating in residential areas, and industrial fuel combustion for processes like steel and paper production. Note that CO$_2$ emissions from industrial processes like cement making, which can be substantial, are not included. Data are sourced from \cite{iea2021greenhouse}.}
    \label{fig:figure1}
\end{figure}

Poland finds itself at a significant crossroads as it navigates the complex path of decarbonization, attempting to reconcile its coal-dependent industrial legacy with the EU's ambitious climate targets. Historically, Poland has been one of the most carbon-intensive economies in Europe \cite{antosiewicz2022distributional}, primarily due to its heavy reliance on coal for electricity generation (Fig. \ref{fig:figure1})—a stark contrast with the EU average. As of 2024, coal-fired power plants contributed to approximately 60\% of Poland's electricity mix \cite{qvist2020retrofit} (Fig.\ref{fig:figure1}), which dropped around 10\% in the past couple of years due to large-scale deployment of renewable energy sources. This dependence still poses a substantial challenge within the framework of the European Green Deal, which aims for climate neutrality by 2050 and requires a fundamental transformation in how energy is produced and consumed.

To facilitate this, the EU's 2030 Climate Target Plan proposes a 55\% net emissions cut from 1990 levels, with an ambitious leap from the previously set 40\% target \cite{rivas2021towards, kulovesi2020assessing}. This target underscores the necessity for an expansive economic transformation across the EU, enforced through mechanisms like the ETS, which extends carbon pricing obligations to all fossil fuel-related emissions, thereby economically incentivizing cleaner practices.

Sector-specific mandates have also been established, necessitating at least a 60\% reduction in emissions from buildings and energy sectors, demanding substantial investments in renewable energy \cite{rivas2021towards, kulovesi2020assessing}. Initiatives like the Renovation Wave aim to modernize existing infrastructure to be more energy-efficient \cite{calvetti2021tech}, while the transport sector is expected to transition towards a 24\% renewable energy share by 2030, emphasizing a shift away from high-carbon technologies \cite{bogdanov2021full}.

The Climate Change Performance Index (CCPI) ranks Poland unfavorably at 55th \cite{burck2019climate}, underscoring its challenges in emissions management, renewable energy adoption, and overall climate policy. While Poland's reliance on biomass \cite{wozniak2021bioeconomy} raises questions about forest sustainability and environmental impacts \cite{raihan2024prioritising}, the country's energy future must pivot towards a more diversified energy mix, prioritizing renewables and increased energy efficiency.

Despite these concerted efforts by the EU, Poland faces hurdles. Positioned 75th on the Energy Transition Index (ETI) (\cite{SINGH2019100382, Juszczak2020}, Poland's pathway reveals a heavy reliance on fossil fuels and a slow pace toward renewable integration. The ETI suggests that without immediate and expansive policy shifts, Poland's achievement of climate neutrality might extend beyond 2056, further challenged by business environments that may deter long-term clean energy investments \cite{Juszczak2020}. However, Poland's potential for implementing CCS technology, particularly in its heavy industries and sectors like steel and chemicals \cite{gajdzik2023process}, is recognized, which could significantly reduce its carbon footprint.

Regarding EU climate policy, the renewed commitment to the Green Deal and its binding resolutions reinforces the imperative for technological transformation as well as social and economic transitions, addressing the just transition for coal-dependent communities in Poland. Initiatives like the EU's Just Transition Fund \cite{moesker2022just, leppanen2022agenda} are crucial in this regard, aiming to mitigate social impacts and promote equitable economic perspectives for those transitioning away from coal.

Poland’s decarbonization narrative must embrace a holistic strategy, integrating technological innovation, economic restructuring, and robust policy frameworks to effectively transition towards a low-carbon economy. Cooperative ventures, such as the Norwegian-Polish CCS Network, could facilitate the development of a cohesive strategy, increasing Poland’s capabilities to meet both its national energy demands and EU-imposed climate objectives.

Poland's substantial geological CO$_2$ storage capacity presents an appealing opportunity for CCS technology deployment, potentially offsetting emissions from its largest polluters. Challenges such as public skepticism, the need for comprehensive regulatory frameworks, and financial investments remain, but engagement with Norway's extensive CCS expertise could prove transformative. Norway has set a precedent, showcasing operational success and public engagement strategies that may serve as valuable blueprints for Poland.

Moving forward, incentivizing investments in renewables, streamlining bureaucratic processes, and deploying CCS at industrial sites are crucial measures. Enhanced public awareness and educational initiatives, alongside governmental transparency in climate strategies, will further solidify Poland’s position in meeting EU climate mandates. The international cooperation could catalyze the transition, turning Poland's decarbonization challenges into opportunities for a sustainable and prosperous future. Poland's engagement in CCS initiatives and collaboration with Norway may present a transformative potential within its climate strategy. Participating in projects like the ECO2CEE (a continuation of the EU CCS Interconnector) \cite{gravaud2023ccus, Orlan2025} can integrate into the broader European CO$_2$ storage infrastructure, ensuring a more resilient and secure emissions containment strategy.

\section{CCS in Poland: A Comprehensive Overview}

Poland is undergoing a significant transition in its CCS landscape, driven by recent legislative reforms and commitments to align with EU climate targets. A pivotal change in policy now permits onshore CO$_2$ storage, a critical legislative shift from previous prohibitions that unlocks Poland's extensive storage potential. Despite these progressive regulations, the realization of full-scale operational CCS projects remains a future goal. Current efforts are concentrated on feasibility studies, pilot projects, and developing a comprehensive CCS strategy. This strategy involves evaluating storage capabilities, expanding transport infrastructure, and fostering public acceptance while leveraging the potential of depleted oil and gas reservoirs and deep saline aquifers for CO$_2$ storage.

Poland's initial forays into CCS were characterized by ambitious projects that encountered numerous challenges. Notable among these were the Bełchatów CCS Demonstration Plant \cite{kapetaki2017highlights, gladysz2020techno} and the Kędzierzyn-Koźle Zero-Emission Power \& Chemical Plant \cite{uliasz2016perspectives}. Launched between 2009 and 2013, the Bełchatów plant integrated carbon capture with a lignite-fired unit, aiming to capture 1.8 million tonnes of CO$_2$ annually using advanced amine-based technologies and dedicated transport pipelines \cite{gladysz2020techno}. Concurrently, the Kędzierzyn-Koźle project sought near-zero emissions goals by converting hard coal into synthesis gas via Integrated Gasification Combined Cycle (IGCC) technology, capturing an impressive 92-93\% of CO$_2$ \cite{uliasz2016perspectives}. This project also explored dual uses for captured CO$_2$, including geological storage in saline aquifers and chemical sequestration in methanol and urea production. However, both projects faced substantial setbacks, ranging from financial constraints to regulatory hurdles hindering their progress.

The contemporary CCS framework in Poland is being reshaped by recent updates to the Geological and Mining Law \cite{uliasz2016perspectives, nagy2022new}, highlighting regulatory and strategic advancements. These legislative changes eliminate restrictions that previously confined CCS to demonstration projects, facilitating smaller projects and enabling enhanced oil recovery (EOR) operations, thereby achieving greater economic efficiency. The formation of the CCUS Working Group underscores Poland's commitment to expanding its CCS capacity. This effort is driven by tailored legislative frameworks and public-private partnerships to transform CCS potential from theoretical concepts into practical, operational applications. These developments signal a strategic shift toward positioning CCS as a core component of Poland's climate action strategy.

\subsection{Carbon capture and storage potential}
Poland's economic transformation over the last 30 years has seen a shift from a planned economy to a more service-oriented model, with GDP nearly tripling \cite{piatkowski2018europe, mrozowska2021challenges}. Yet, the nation remains amongst the EU's most carbon-intensive economies, emitting approximately 800 grams of CO$_2$e per euro of GDP \cite{McKinsey2020}. This high carbon intensity reflects the dominance of coal and lignite in Poland's energy mix, despite increasing shares of renewables and industrial energy efficiency improvements \cite{McKinsey2020}. Achieving net-zero emissions by 2050 requires a fourfold increase in the rate of decarbonization over the next decade compared to the past thirty years \cite{kochanek2021evaluation, McKinsey2020}. Large-scale infrastructure renewal focusing on zero-carbon technologies provides an opportunity to leverage Poland’s forest carbon sinks to offset emissions from hard-to-abate sectors such as agriculture.

\begin{figure}
    \centering
    \includegraphics[width=0.9\textwidth]{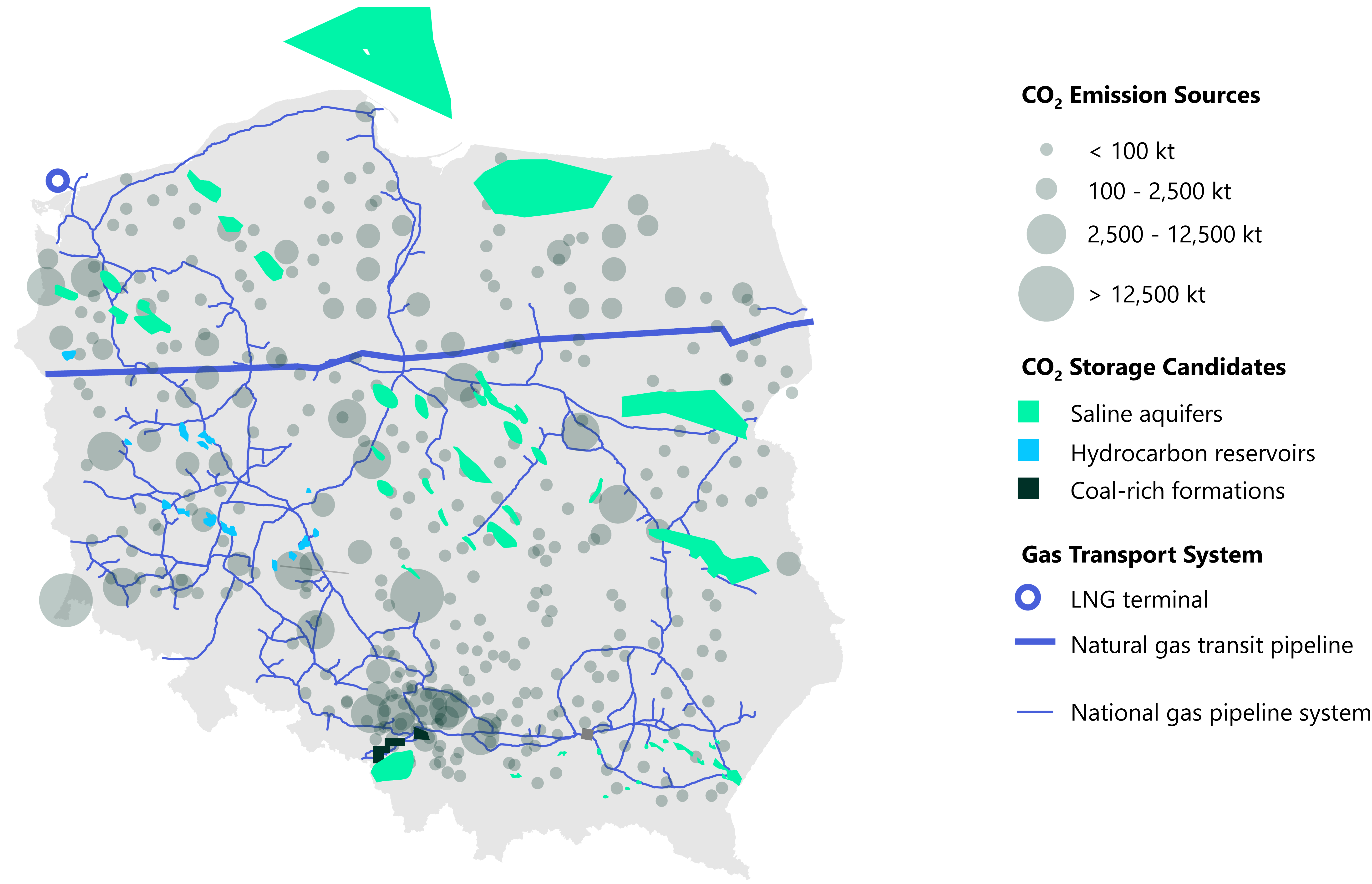}
    \caption{Map of major CO$_2$ emission point sources and potential geological storage candidates in Poland. This map categorizes emitters into four groups and highlights potential sites for CO$_2$ sequestration. Poland's storage capacity (rough static computations) is preliminarily estimated at 15 gigatonnes (Gt) of CO$_2$, distributed as follows: 14.3 Gt in saline aquifers, 1.0 Gt in hydrocarbon reservoirs, and 0.1 Gt in coal fields. The data and figures are adapted from the Polish Geological Institute's "Interactive Atlas Presenting Possibilities of CO$_2$ Geological Storage in Poland" in addition to \cite{McKinsey2020, iea2021greenhouse}.}
    \label{fig:figure2}
\end{figure}

Poland’s greenhouse gas emission profiles reveal key contributors \cite{McKinsey2020, owid-CO2-dataset-sources, crippa2024insights, iea2021greenhouse} (Fig.\ref{fig:figure1}): power and heat generation accounts for 38\% (156 MtCO$_2$e) of emissions, with the majority stemming from coal and lignite combustion. Industry contributes 22\% (91 MtCO$_2$e), dominated by subsectors including fuel manufacturing and chemicals. Transportation contributes 15\% (63 MtCO$_2$e) and buildings account for 11\% (46 MtCO$_2$e), primarily from coal and gas usage for heating. Agriculture, contributing 11\% (44 MtCO$_2$e), is characterized by non-CO$_2$ emissions, with waste and miscellaneous sectors adding an additional 3\%. Notably, Poland benefits from -34 MtCO$_2$e of negative emissions \cite{EULULUCF}, predominantly via carbon removal in the Land Use, Land-Use Change, and Forestry (LULUCF) sector, primarily through existing forests and increased forest coverage.

As shown in Figure \ref{fig:figure2}, the geological resources of Poland promise significant CO$_2$ storage potential \cite{lubon2020co, urych2022numerical}, offering diverse formations: saline aquifers possess an estimated capacity of 14,3 Giga tonnes (Gt); depleted hydrocarbon reservoirs add 1,0 Gt, potentially 0.1 Gt storage in coalbed fields. There is a suggested possibility of additional offshore pore space in the Polish economic zone contributing another 0.8 Gt. The total estimated static capacity of CO$_2$ sequestration suggests potential storage for the foreseeable future (approximately 50-75 years) of Poland's current ETS emissions levels. Saline aquifers, particularly in regions like Bełchatów, Warsaw, and Kujawy, hold the most significant potential. However, detailed assessments are necessary due to potential conflicts and to confirm the capacity and sealing properties in formations such as the Palaeozoic basins.

Enhanced oil recovery (EOR) capitalizes on established infrastructure and geological knowledge to improve the utility of existing oil and gas reservoirs for CO$_2$ storage. Meanwhile, unexploited deep coal seams present opportunities for CO$_2$ storage paired with methane extraction, although safety, capacity, and economic factors require careful evaluation. Robust Measurement, Monitoring, and Verification (MMV) protocols are needed to ensure safe storage.

\subsection{Past projects and present initiatives}
The Bełchatów CCS plant \cite{kapetaki2017highlights, gladysz2020techno}, although not realized, showcased the potential integration of full carbon-capture lifecycle processes. Tauron’s mobile pilot projects at Łaziska and Łagisza illustrated the technical potential of amine-based and pressure-swing adsorption technologies, revealing the feasibility of deploying mobile capture systems across different sites \cite{miu2021assessment, sliwinska2022carbon}. Further, projects like the Borzęcin enhanced gas recovery pilot validated CO$_2$ utilization for natural gas purification, reinforcing the applicability of CCS in resource extraction \cite{warnecki2021study, knapik2020chemistry}. The RECOPOL project \cite{checko2020research} further demonstrated the feasibility of CO$_2$ storage in deep, unmineable coal seams within the Upper Silesian Coal Basin, providing valuable data on storage parameters. Meanwhile, innovative projects like the CO$_2$ Methanation System at the Łaziska Power Station illustrate synergies between CO$_2$ capture and alternative fuel production pathways, particularly SNG \cite{chwola2020pilot}.

The recent ECO2CEE (former Poland-EU CCS Interconnector) at the Port of Gdansk \cite{gravaud2023ccus, Orlan2025} (Fig.\ref{fig:figure2}), which will establish a hub for CO$_2$ import-export utilizing multi-modal transport connectivity with European networks, may accommodate up to 8.7 million tonnes of annual CO$_2$ transport by 2035. The ECO$_2$CEE project solidifies the infrastructure, leveraging multiple transport modes, including rail, road, waterway, and pipeline options for optimized CO$_2$ distribution \cite{gravaud2023ccus, wojnicki2023perspectives}. The Go4ECOPlanet initiative aims to decarbonize Holcim's cement processing in Kujawy via advanced CryoCap capture systems, targeting total decarbonization by transporting and storing captured CO$_2$ offshore \cite{strojny2024preliminary}.

The latest strategy of the ORLEN Group, unveiled in January 2025 \cite{Orlan20252}, highlights the significance of CCS and CCU technologies across the downstream, upstream, and supply value chains. By 2035, the proposed carbon management services aim to encompass the terminal in Gdansk and achieve a transport and storage capacity of 4 million tons of CO$_2$ annually. This target includes 1.1 million tons per annum (Mtpa) for ORLEN Group's own needs, 1 Mtpa allocated to the Go4ECOPlant initiative for external capacity, and 1.9 Mtpa designated for commercial capacities.

\subsection{Challenges and legislative considerations}
Even amid recent legislative progress, Poland's CCS initiatives face continued regulatory challenges. The EU's vigorous decarbonization efforts reflected, for instance, in amendments to the EU ETS Directive, have tightened emissions allowances, compelling industries to adopt more advanced technologies. However, CCS remains a crucial option for sectors with emissions inherent to production processes, such as cement and chemical manufacturing.

At the EU level, the legal framework for CCS is established by the directive on the geological storage of carbon dioxide, which is integrated into Polish law through the Geological and Mining Law. This law details the procedures for obtaining licenses for CO$_2$ storage—granted after consultation with the European Commission—and outlines the obligations and responsibilities of businesses, including site planning and environmental impacts. However, critical gaps remain, as current legislation largely overlooks the capture phase and addresses CO$_2$ transportation only through general Energy Law.

Historically, Polish regulations restricted CCS projects to demonstration purposes only, due to prior safety concerns, which effectively limited commercial development. Despite this precaution, no demonstration projects materialized, highlighting the restrictive nature of past regulations. Recent amendments to the Geological and Mining Law, however, have opened the field to commercial CCS projects by allowing smaller CO$_2$ storage initiatives and proposing economic incentives through EOR applications.

The EU’s Net Zero Industry Act, requiring member states to achieve significant CO$_2$ injection capabilities by 2030, imposes new momentum on Poland’s CCS efforts. This Act promotes streamlined administrative processes, establishing points of contact for CCS permitting to minimize approval timeframes to 18 months. Aligning with such overarching EU mandates can secure Poland’s access to essential financial and policy support, driving long-term development.

Despite these advancements, the public perception of CCS remains a formidable barrier, demanding effective stakeholder engagement. Building trust necessitates a concerted effort to transparently communicate the environmental and economic benefits of CCS and proactively address safety concerns. Engaging communities early and incorporating their input into planning phases can mitigate opposition and foster a collaborative environment, which is essential for successful project deployment.

Moreover, the EU’s broader CCS framework supports transnational infrastructure projects, such as cross-border CO$_2$ networks prioritized under Regulation TEN-E. Such initiatives facilitate Poland’s integration into networked European CCS systems, enhancing both its technical capacity and economic viability.

\subsection{Outlook and strategic steps for deployment}
Despite acknowledging CCS in the National Energy and Climate Plan (NECP) \cite{NECP} and Energy Policy of Poland until 2040 (EPP2040) \cite{EPP2040}, CCS does not yet play a central role in Polish national planning. The NECP primarily frames CCS within the research and innovation sphere, citing implementation challenges noted in 2017 \cite{NECP}. EPP2040 mentions CCS as a tool for reducing coal dependence, highlighting that Polish coal plants are CCS-ready, but lacks a detailed strategy with specific targets or timelines \cite{EPP2040}. Profitability issues also loom, as EPP2040 suggests CCS becomes viable only when CO$_2$ allowance prices exceed EUR 50 per tonne.

The Polish Hydrogen Strategy \cite{PolHyd} offers a glimpse of CCS potential in clean hydrogen production from natural gas by 2030, pending public consultation outcomes. Additionally, the non-binding social agreement with mining unions outlines a gradual coal phase-out by 2049, incorporating CCU/CCS projects like coal gasification with Integrated Gasification Combined Cycle (IGCC) technology and CO$_2$ capture. However, uncertainties in financing and realization timelines remain significant hurdles.

Despite these planning document gaps, legislative frameworks like the Geological and Mining Law align with the European CCS Directive, providing a base for progressing beyond demonstration purposes. The lack of funding specificity and dedicated support remains a notable barrier, notwithstanding EU support plans such as the Connecting Europe Facility intended for 2021-2025.

Within Poland's strategic landscape, the newly established CCUS Working Group serves a critical function by facilitating collaboration and tackling the primary challenges associated with deployment. The group is actively engaged in assessing the feasibility of incorporating CCS as a comprehensive strategy tailored for various industrial sectors. These efforts signify a strategic move towards industrial decarbonization through the creation of clusters, optimizing operational efficiency and cost-effectiveness by addressing the needs of multiple emitters simultaneously.

As the industrial and legislative environment evolves, steps to integrate CCS into the National Energy and Climate Plan are crucial for coherence with EU mandates such as the Net Zero Industry Act. This Act mandates significant CO$_2$ injection capabilities by 2030 and introduces streamlined administrative processes, including designated points of contact for CCS permitting to expedite project approvals. Aligning national policies with EU objectives is vital for leveraging financial and regulatory support. Looking forward, Poland should anticipate an expansive role for CCS by 2040, utilizing developed transport networks and regional hubs. Effective implementation of specialized projects like CCS deployment in hard-to-abate industry and potential CCS-ready power stations will be pivotal in contributing to Poland's climate neutrality aspirations. The Ministries of Climate and Environment, alongside other Ministries, are instrumental in charting this path, supported by industrial stakeholders and research institutions.

\section{Norway's CCS Endeavors Encapsulated}

Norway stands at the forefront of global CCS R\&D and industrial endeavors, leveraging a rich history of active involvement and public engagement. Operational projects that began in the mid-1990s have established Norway as a benchmark for CCS activities worldwide. This commitment is reinforced by a robust legal framework, substantial investments from both public and private sectors, and proactive research and development initiatives. Presently, Norway is enhancing its CCS infrastructure to accommodate CO$_2$ storage from domestic and international sources through the incorporation of available technologies and subsurface resources in the Norwegian Continental Shelf (NCS).

\begin{figure}
    \centering
    \includegraphics[width=0.7\textwidth]{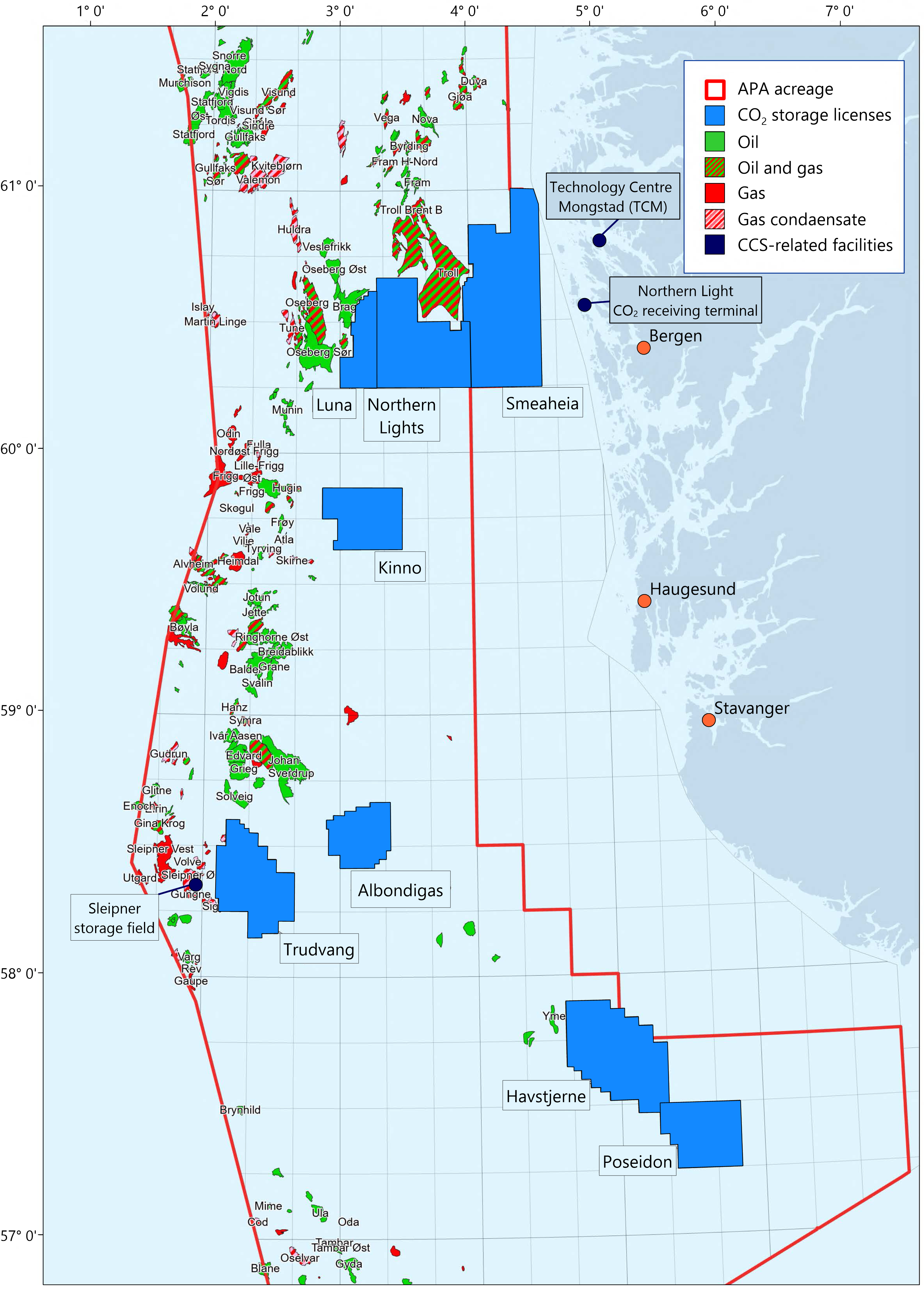}
    \caption{Map of the Norwegian Continental Shelf (NCS) showing petroleum fields offshore Norway and eight CO$_2$ storage licenses in the North Sea. It also presents active CCS facilities, such as the Sleipner storage site. Note that the Polaris license and the Snøhvit geological storage site in the Barents Sea are not included within the scope of this map. The data and map elements are adapted from \cite{SD2025, Eq2025}.}
    \label{fig:figuremap}
\end{figure}

Norway's climate targets include a 40\% reduction in greenhouse gas emissions by 2030 compared to 1990 levels and climate neutrality by 2050 \cite{NorClim2130, malka2023energy}. The engagement in CCS was reinforced when the integration of EU CCS Directive (2009/31/EC) into Norwegian law. However, Norway's experience with CCS predates this directive. The Sleipner project, operational since 1996, serves as a key example, safely storing over 20 million tonnes of CO$_2$ to date \cite{furre2019building, furre2024sleipner} ((Fig.\ref{fig:figuremap})). Similarly, the Snøhvit project, initiated in 2008, captures and stores approximately 0.7 million tonnes of CO$_2$ annually \cite{furre2019building, ringrose2020store}.

A critical component of Norway's CCS strategy is the Northern Lights / Longship project, which aims to build an open-access CO$_2$ transport and storage network \cite{wang2024status, al2021review}. This involves a significant financial commitment, approximately NOK 25.1 billion (EUR 2.5 billion) \cite{trupp2022risk} to facilitate CO$_2$ capture, transport, and storage for industrial emitters nationally and across Europe. The NCS strengthens Norway's CCS readiness, offering an estimated storage capacity of 70-80 Gt in deep saline aquifers and depleted reservoirs \cite{halland2019offshore, riis2014co2}. The country also benefits from well-defined regulatory frameworks \cite{santos2021regulatory} that support CCS from exploration to long-term monitoring.

There is growing interest among companies in CO$_2$ storage on the NCS, supported by government initiatives aimed at fostering profitable carbon sequestration ventures. Companies possessing the requisite expertise and well-defined business strategies are encouraged to apply to the Ministry of Energy for permits specifically tailored to their operational needs, as shown in Figure \ref{fig:figuremap}. By October 2024, the Ministry had issued eleven permits under CO$_2$ storage regulations, ten of which are located in the North Sea, and one in the Barents Sea. Furthermore, in June 2024, the announcement of three new storage areas expanded the existing zones, all positioned in the North Sea.

In addition to its substantial storage capacity in NCS, Norway offers a robust infrastructure for CCS operations. Existing offshore infrastructure, originally developed for oil and gas activities, presents opportunities for repurposing into CO$_2$ transport systems. The Long Ship project further enhances this capability by establishing dedicated CO$_2$ transport networks, utilizing custom-designed CO$_2$-transport ships and a receiving terminal anticipated to handle 1.5 million tonnes of CO$_2$ annually by 2024. The project's future phases aim to expand this capacity beyond 5 million tonnes. In this context, Norway is considering CO$_2$ shipping as a viable transportation strategy, thereby augmenting the flexibility of its infrastructure and facilitating international CO$_2$ reception via specialized carriers.

Furthermore, clarifying long-term liability for CO$_2$ storage sites is crucial for attracting investment and ensuring safety. Although mechanisms for liability transfer to the state exist, further clarity is required, alongside international cooperation to facilitate cross-border CCS projects like Northern Lights. Integration within the EU ETS needs refinement to enhance long-term investment certainty, ensuring captured CO$_2$ is recognized and rewarded within the framework.

Several key factors contribute to Norway's remarkable success in advancing CCS initiatives. The country benefits from robust political support, which ensures policy continuity and stability, which are crucial for long-term project planning. This political backing is complemented by proactive industry engagement, particularly from the oil and gas sector, where companies such as Equinor play pivotal roles in driving technological innovation and optimizing project execution. Norway's success is further bolstered by the contributions of its dedicated environment-friendly energy research national centers (FME Centers) led by universities and research institutions, which spearhead groundbreaking research and innovation in the CCS field. Moreover, public awareness and acceptance are enhanced through strategic communication efforts \cite{anders2024public, merk2023carbon}, a vital component in fostering public confidence and support for ongoing CCS projects. These factors collectively create a synergy that positions Norway as a leader in implementing effective and sustainable CCS solutions.

Norway's dedication to CCS presents significant opportunities for the country on the global stage. By establishing itself as a hub for CCS technology and services, Norway can export its expertise worldwide, offering comprehensive storage solutions and expert consulting services. This strategic position can drive economic diversification, particularly through the creation of new jobs. Additionally, integrating CCS into hydrogen production from natural gas aligns seamlessly with Norway's ambition to become a leader in the hydrogen economy \cite{damman2021hybrid, larsgaard2024innovation, cheng2024competitive}, thereby supporting decarbonization efforts across various sectors, including transportation and industry.

Looking to the future, Norway aims to solidify its position as a global leader in CCS, with aspirations to have multiple large-scale CCS projects operational by 2040, including the Northern Lights initiative. Continuous development of CO$_2$ transport networks remains crucial for both domestic and international demands, forming a flexible, comprehensive system that integrates seamlessly with national climate policies. By 2040, Norway envisions itself as a major CCS hub, extending its storage services globally and achieving widespread industrial adoption across various sectors. This transition is anticipated to involve close industry collaboration and ongoing innovation, contributing significantly to Norway's national targets and international net-zero objectives.

\section{Bilateral Network As A Decarbonization Bridge}

The urgent need to mitigate climate change necessitates the development of innovative, cross-border solutions. A notable initiative in this context could be the collaborative model between Norway and Poland to establish a CCS network. This partnership leveraged Norway’s pioneering expertise in CCS and capitalized on Poland’s substantial interest in carbon removal technologies, its extensive geological storage potential, and its significant industrial emission profiles. This strategic alliance transcends bilateral cooperation, offering a scalable and comprehensive blueprint for pan-European CCS development that aligns with the European Union’s climate neutrality targets.

\subsection{Synergic collaboration components}

The synergistic collaboration between Norway and Poland holds the promise of significantly advancing Poland’s transition to a low-carbon economy. By joining forces, the two nations can address the technical and operational challenges that impede widespread CCS adoption and deployment. For Poland, this cooperation accelerates its decarbonization process, allowing the country to harness its extensive geological storage potential and industrial capabilities. Meanwhile, for Norway, it represents a strategic opportunity to further broaden its CCS activities, reinforcing its leading position in the development of diverse value chain components, including industrial applications, R\&D initiatives, and socio-economic as well as legal-policy frameworks.

Given the multifaceted dynamics of excellence and challenges inherent at both ends of this cooperation and based on an analysis of near-term and long-term perspectives outlined in previous sections, the overarching aim of this initiative was set to establish a sustainable bilateral network. This network would connect Norwegian and Polish entities with the long-term vision of fostering collaborations across the entire CCS value chain. This involves bridging the industry, business, and research sectors to expedite CCS deployment initiatives. Key institutions such as AGH University of Krakow, the University of Oslo, and Norwegian Energy Partners were integral to this endeavor. The project team has outlined activities with well-defined objectives: aligning needs with expertise and implementation pathways, facilitating direct contact and matchmaking among stakeholders, and promoting competence-building and infrastructure development, with a strong emphasis on proactive political and public engagement. To achieve these goals, a series of strategic initiatives have been outlined, as depicted in Figure \ref{fig:figure3}:

\begin{itemize}

\item \textit{Business-to-Business (B2B) networking:} This initiative focused on fostering connections among key stakeholders in the CCS sectors of both Poland and Norway, encompassing industry leaders, businesses, and research and development (R\&D) entities. A fundamental element of B2B networking is the alignment of stakeholders, which establishes a cohesive vision critical to efficient strategy execution. This initiative creates platforms for stakeholders to engage in discussions around shared goals, challenges, and opportunities in CCS implementation. Here, cross-border partnerships play an instrumental role by enabling the exchange of expertise, resources, and insights that result in complementary strengths and potential joint ventures.

The B2B networking provided a platform for stakeholders to engage in discussions around shared goals, challenges, and opportunities in CCS implementation. These interactions facilitated the identification of complementary strengths and resources, paving the way for potential joint ventures and projects. Moreover, this initiative prioritized the dissemination of innovative strategies and best practices, aiming to accelerate deployment. By leveraging the collective experience and capabilities of Polish and Norwegian entities, the B2B networking initiative significantly contributed to reinforcing the interest and competence necessary for joint CCS ventures.

\item \textit{CCS educational program:} It consists of two core components designed to enhance knowledge and foster collaboration:
\begin{itemize}
    \item A. Industrial site visits, where delegations from the Polish industry, research and development (R\&D), and policymaking sectors participated in immersive visits to Norwegian CCS facilities. These visits provided valuable insights into CCS implementation processes and facilitated active stakeholder engagement by allowing participants to observe ongoing deployment efforts in Norway firsthand.
    \item B. CCS summer school in Poland, an interdisciplinary program aimed at raising awareness and equipping the younger generation and early career professionals with essential knowledge and skills. This initiative brought together technical, societal, and policymaking experts to deliver comprehensive training from both academic pursuits and industry-oriented activities. 
\end{itemize}

Crucially, the Norwegian industry and R\&D sector were actively engaged in both components, sharing the latest developments, state-of-the-art technologies, and lessons learned from ongoing industrial projects. This collaboration provided participants with tangible advanced knowledge across the capture, transport, and storage spectrum.

\begin{figure}
    \centering	
    \includegraphics[width=0.9\textwidth]{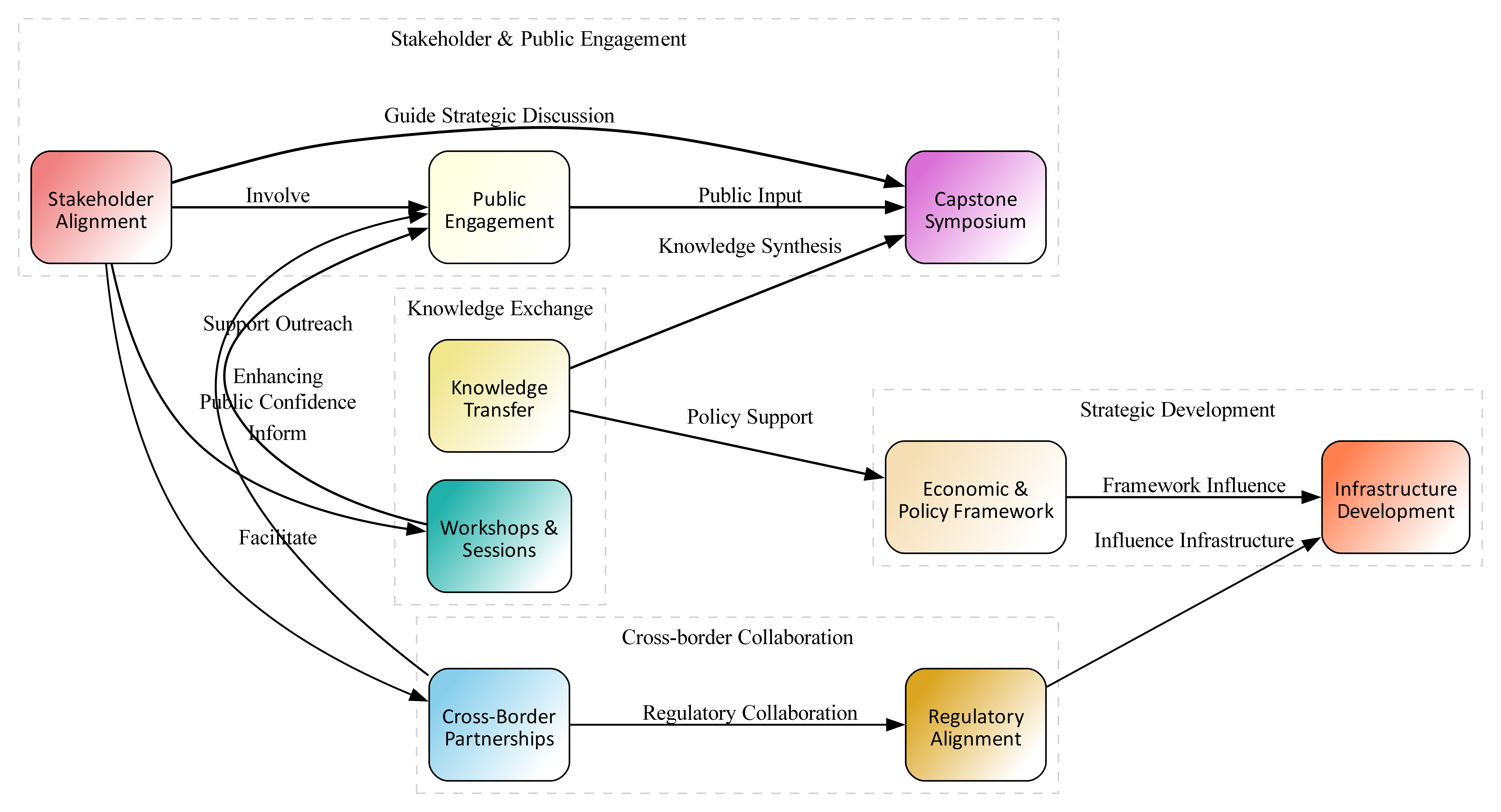}
    \caption{Synergic collaboration components depicted as an interconnected array designed to drive collaborative CCS initiatives between Norway and Poland. Nodes represent pivotal elements, and clusters denote thematic domains. The edges indicate directional influence among components, highlighting critical pathways for strategy alignment within the collaborative network. This graph provides a dynamic representation of the operational synergy essential for effective bilateral cross-border CCS cooperation.}
    \label{fig:figure3}
\end{figure}

\item \textit{Workshops, brainstorming gatherings, and capstone symposium:} The series of workshops, brainstorming gatherings, and the capstone symposium were strategically designed to refine competencies and align interests among participants, thereby promoting business development and enhancing research cooperation. These interactive events served as incubators for collaborative action plans, providing attendees with the opportunity to explore cutting-edge advancements, discuss challenges, and identify potential solutions tailored to their needs. Brainstorming gatherings specifically encouraged open dialogue and collaboration among diverse stakeholders, generating problem-solving approaches and fostering a sense of community within the network.

The Capstone CCS Symposium was a pinnacle event that brought together active stakeholders from Poland, Norway, and beyond to serve as a platform for bilateral cooperation. This symposium was dedicated to the comprehensive exploration of CCS deployment strategies, offering a forum for participants to discuss findings and share best practices for moving forward in light of policy implications. The event not only strengthened existing partnerships but also created new avenues for collaboration, aligning goals and resources toward joint initiatives. By connecting a broad spectrum of professionals, the initiative cultivated a dynamic and supportive ecosystem essential for expediting activities in Poland via this bilateral forum. 

\item \textit{Public relations campaigns for raising awareness and gauging public opinion in Poland:} The public relations campaigns were a comprehensive initiative designed to elevate public awareness and understanding of the CCS value chain as a vital tool for sustainable climate change mitigation. The campaigns aimed to inform the general public about the significance of CCS technologies and their role in reducing carbon emissions, thereby supporting bilateral and international cooperation. The campaign employed a multifaceted approach, utilizing diverse communication channels to reach a broad audience. Engaging content was disseminated through online and social media outlets, ensuring widespread visibility across various demographics. 

The public opinion surveys were conducted in two comprehensive phases, with the primary objective of evaluating Polish individuals' attitudes and perceptions regarding climate change, emission reduction technologies, and their potential implementation in Poland. In the initial phase, the survey focused on establishing a baseline understanding of public awareness about CCS and gauging preliminary attitudes toward it. The second phase, implemented following targeted informational campaigns and educational initiatives, aimed to assess changes in perception and knowledge levels, providing insights into the effectiveness of these interventions.

The surveys measured public awareness and interest in CCS and explored deeper sentiments regarding environmental policies, technological acceptance, and trust in stakeholders involved in such initiatives. By analyzing shifts in public opinion over time, the studies provided valuable data on potential barriers to CCS adoption, highlighted opportunities for educational outreach, and informed stakeholders about public readiness and support for further technological integration. This information proved critical in shaping future communication strategies and aligning public engagement efforts with the overarching goal of sustainable climate action.
\end{itemize}

The synergic collaboration pathways (Fig. \ref{fig:figure3}), visualize the designed framework underpinning the collaborative efforts between Norway and Poland. The nodes within the graph represent critical components such as stakeholder alignment, public engagement, and workshops, capturing the fundamental elements guiding this collaboration. By interlinking nodes via weighted edges, the visualization underscores the dynamic interplay and influence each component exerts within the synergistic framework—indicating pathways like knowledge transfer and economic policy support, which collectively drive cohesive action. Clusters are proposed to group related components into thematic domains, such as Stakeholder \& Public Engagement and Cross-border Collaboration, which illustrate the broad-based integration of efforts required to meet ambitious decarbonization goals. The Capstone Symposium node, epitomizes the culmination of strategic discussions, reinforcing the collective-driven approach to CCS initiatives. Through these elements, the visualization encapsulates the seamless convergence of business, education, and innovation strategies deployed within this bilateral alliance, thereby providing an insightful overview of how Norway and Poland may foster operational excellence and meet the challenges of today’s low-carbon initiatives.

\subsection{Stakeholder Alignment and Value Proposition}

The designed collaboration components were focused on applying the Unique Value Proposition (UVP) and/or Unique Selling Point (USP) frameworks to harmonize Norwegian competencies with the particular needs and challenges faced by Polish stakeholders. By crafting tailored solutions aligned specifically with these needs, the Norwegian-Polish CCS Network's core team sought to establish robust, fit-for-purpose collaborative partnerships among interested parties. Central to this approach was the identification of distinctive points of differentiation, which underscored the added value that Norwegian expertise could bring to Polish CCS stakeholders. This strategic guidance was vital in ensuring purposeful engagement, ultimately fostering a conducive environment for potential project development and execution.

As a part of this strategic alignment, as mentioned above, targeted workshops were conducted in both Norway and Poland, which were structured around three primary themes: carbon capture for industrial applications, carbon storage solutions, and research and development (R\&D) cooperation. These workshops extended to encompass joint sessions that addressed the multifaceted dimensions of legal, economic, financial, policy, and societal considerations. Such a holistic approach laid down a robust foundational framework for potential future joint CCS deployments in Poland, substantially supported by active Norwegian participation. In addition, the initiative explored sustainable legal and regulatory frameworks, carefully balancing economic viability, social acceptance, and environmental sustainability.

Furthermore, proactive stakeholder engagement from both countries was facilitated through a series of meetings, brainstorming sessions, and open discussions. These interactions were pivotal in fostering open dialogue, allowing for the identification of opportunities, challenges, and potential obstructions. A retrospective evaluation of previous collaborative efforts provided valuable insights into lessons learned while uncovering the core motivations of stakeholders. This reflective process was crucial in garnering individual and collective support for future projects, thereby creating a pathway for streamlined collaboration.

A significant focus was placed on the legislative landscape, with regulatory frameworks in Poland identified as a primary area needing clarification and better harmonization with EU directives. Understanding and addressing these legal impediments was crucial, as they represented a substantial barrier in the Polish market for CCS implementation. By clarifying and smoothing the path for collaborative partners within both the Norwegian and Polish contexts, the initiative underscored the importance of aligning legislative initiatives to enhance the feasibility and success of bilateral cooperation.

\subsection{Demand Drivers and Opportunities Identification}

Through the comprehensive activities conducted under the collaboration components outlined in Section 4.1, the core team engaged in extensive analysis with network participants, complemented by a detailed evaluation of Polish national documents and related reports. This effort aimed to elucidate the diverse demand drivers for CCS/CCUS deployment in Poland while simultaneously capturing the perspectives of Norwegian industrial and R\&D stakeholders.

A primary driver identified is civil society's expectation of a transition towards a future characterized by net-zero emissions. Government policy and regulations are increasingly reflecting this societal demand, thereby strengthening the business case for investing in CCS technologies. CCS is recognized as a vital instrument for reducing CO$_2$ emission exposure in specific industrial sectors, presenting a dual opportunity to minimize greenhouse gases while fostering a burgeoning industry. It is positioned within scenarios as one of the most efficient pathways to achieve net-zero targets, ensuring a just transition for industrial and highly carbon-dependent communities.

Simultaneously, as reviewed earlier, significant legislative progress has been achieved, most notably through the amendment to the Geological and Mining Law. This amendment now permits full-scale commercial onshore and offshore CO$_2$ storage activities in Poland and encourages the geological extraction of hydrocarbons in tandem with underground carbon dioxide storage. This legislative evolution creates an enabling environment that catalyzes investment and development within the sector.

From a strategic perspective, Poland's CCS landscape presents a multifaceted spectrum of opportunities and threats across several critical dimensions. Economically, there is substantial potential for growth and investment stimulation; however, quantitative financial feasibility challenges pose a risk that necessitates careful consideration and strategic financial planning. Socially, there is considerable promise in garnering public support and fostering community resilience through emission-reduction technologies. Nevertheless, a significant threat is posed by potential public resistance, driven by social concerns and misinformation, making proactive communication and education initiatives essential. This reflects the critical importance of public engagement strategies and infrastructure as outlined in the Advanced Demand Drivers and Opportunities diagram, which emphasizes how stakeholder alignment and public outreach can enhance trust and consensus-building.

Technologically, the field offers immense opportunities for business development, yet significant technological hurdles could impede progress without focused R\&D efforts. The broad potential for CCS implementation spans diverse industries, each presenting remarkable opportunities for decarbonization; however, sector-specific strategies are crucial to address challenges and ensure feasibility and successful adaptation. The diagram also highlights the role of technological development and knowledge transfer in translating advances into practical, scalable solutions, underscoring the need for continuous innovation and education to overcome technical barriers.

In terms of infrastructure, Poland benefits from foundational elements as an initial layer of opportunity. Nonetheless, substantial infrastructural advancements are imperative to bridge existing gaps and ensure robust operational capabilities. Legal frameworks, while evolving, continue to experience growing pains, emphasizing the need for regulatory alignment as depicted in the diagram, which amplifies the interplay between strategic regulatory frameworks and infrastructure growth.

\begin{figure}
    \centering
    \includegraphics[width=0.9\textwidth]{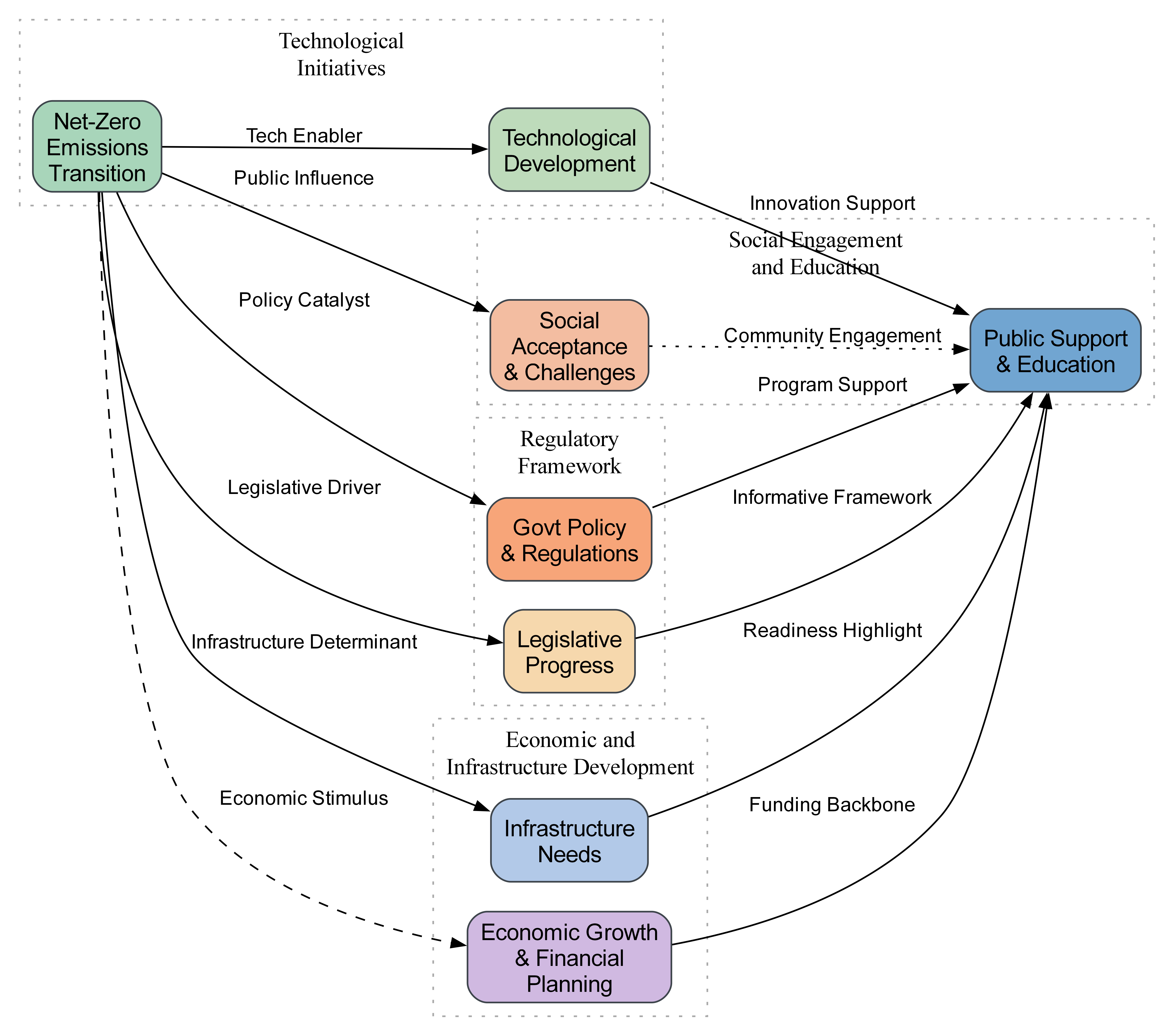}
    \caption{Interplay of governing factors and opportunities influencing the CCS deployment in Poland. Nodes represent key elements required to achieve emission reduction goals. Directed edges illustrate the influences and dependencies among these factors, forming a comprehensive framework to navigate complexities in adaptation and governance. This visualization highlights critical leverage points, policy implications, and strategies for effective stakeholder engagement, serving as a pivotal tool for constructing a roadmap for bilateral Norwegian-Polish cooperation and advancing CCS deployment efforts in Poland.}
    \label{fig:figure4}
\end{figure}

Figure \ref{fig:figure4} effectively visualizes the complex interplay of factors influencing the successful deployment of CCS technology in Poland through a directed graph. Each node in the graph represents a key factor, while directed edges signify their influence on one another, highlighting the multifaceted and interconnected nature of these relationships. Central to this graph is the ``Net-Zero Emissions Transition,'' positioned as the core driver, which underscores the overarching imperative to decarbonize the Polish economy, exerting influence on all other factors. The graph accentuates the critical importance of strong government policies and regulatory frameworks, technological advancements, and social and public dimensions. It also highlights economic considerations, such as the need for infrastructure development, financial planning, and incentives to attract investment, ensuring the economic viability of CCS projects.

Given Poland's reliance on coal for energy production, CCS technology presents a vital pathway toward decarbonization (Figure 2). By visualizing these interdependencies, the plot aids policymakers, researchers, and stakeholders in understanding the challenges and opportunities of deploying CCS. This framework serves to identify critical leverage points, inform policy decisions, prioritize research, and enhance stakeholder engagement to support this essential technology.

The bilateral collaboration identified critical needs for developing a coherent outlook for implementation and scaled-up deployment. It concentrated on four pivotal areas: scaling up research and development activities and enhancing national knowledge and expertise; establishing and refining policies, standards, and regulations; engaging stakeholders to promote cooperation and knowledge diffusion; and addressing social aspects to garner public support. These facets could be framed within two strategic time horizons: the short-term (up to 2030) and the long-term (up to 2040).

The reliance on manufacturing sectors underscores a dependency on fossil fuels for energy production. Within this context and alongside sometimes ambiguous deadlines for emissions reductions, CCS's role in the energy sector remains a viable consideration. Although transportation methods from emitters to storage theoretically exist, the conspicuous absence of comprehensive CO$_2$ transportation infrastructure presents a challenge. Among regional nations, Poland exhibits significant potential for geological CO$_2$ storage (ranking third after Ukraine and Romania). Nevertheless, continued research is essential to refine storage potential estimations, focusing on aspects of storage site selection workflows, transitioning from static to dynamic evaluations of storage capacity, and devising detailed derisking and MMV plans. Poland's CCS landscape, supported by a history of research and pilot testing, stands to benefit from enhanced expertise through bilateral cooperation.

The regulatory environment remains nascent, with numerous aspects yet to provide assured pathways for CCS deployment. While funding support is accessible through European Union channels, including Programs of Common Interest favoring large-scale regional infrastructural endeavors, stakeholder hesitations persist. These apprehensions stem from the high costs, unclear government support, financing difficulties, and complex administrative protocols involved in implementation. A broader challenge is the lack of public and institutional awareness regarding CCS technologies, which the Advanced Demand Drivers and Opportunities diagram seeks to address through its emphasis on strategic public outreach and policy integration frameworks.

Addressing these gaps, particularly in the short term, mandates support for companies and emitters in developing and implementing strategies for decarbonization pathways and regional hubs. This includes assisting them in applying for financial support, such as the Innovation Fund, and facilitating general project development. Given the rapid expansion of the global market, stakeholders must adapt swiftly, necessitating support for know-how and knowledge transfer in organizing the CCUS value chain. Herein lies the effective potential of Norwegian expertise and excellence.

\section{Public Opinion towards CCS in Poland}

Two phases of the public survey were carried out to evaluate the attitudes and opinions of Polish individuals regarding CCS and its implementation prospects in Poland. The study sampled 1,002 Polish adults using computer-assisted web interviews via an Internet-based platform. This representative sample considered demographic factors such as gender, age, town size, and regional distribution.

\begin{figure}
    \centering
    \includegraphics[width=0.9\textwidth]{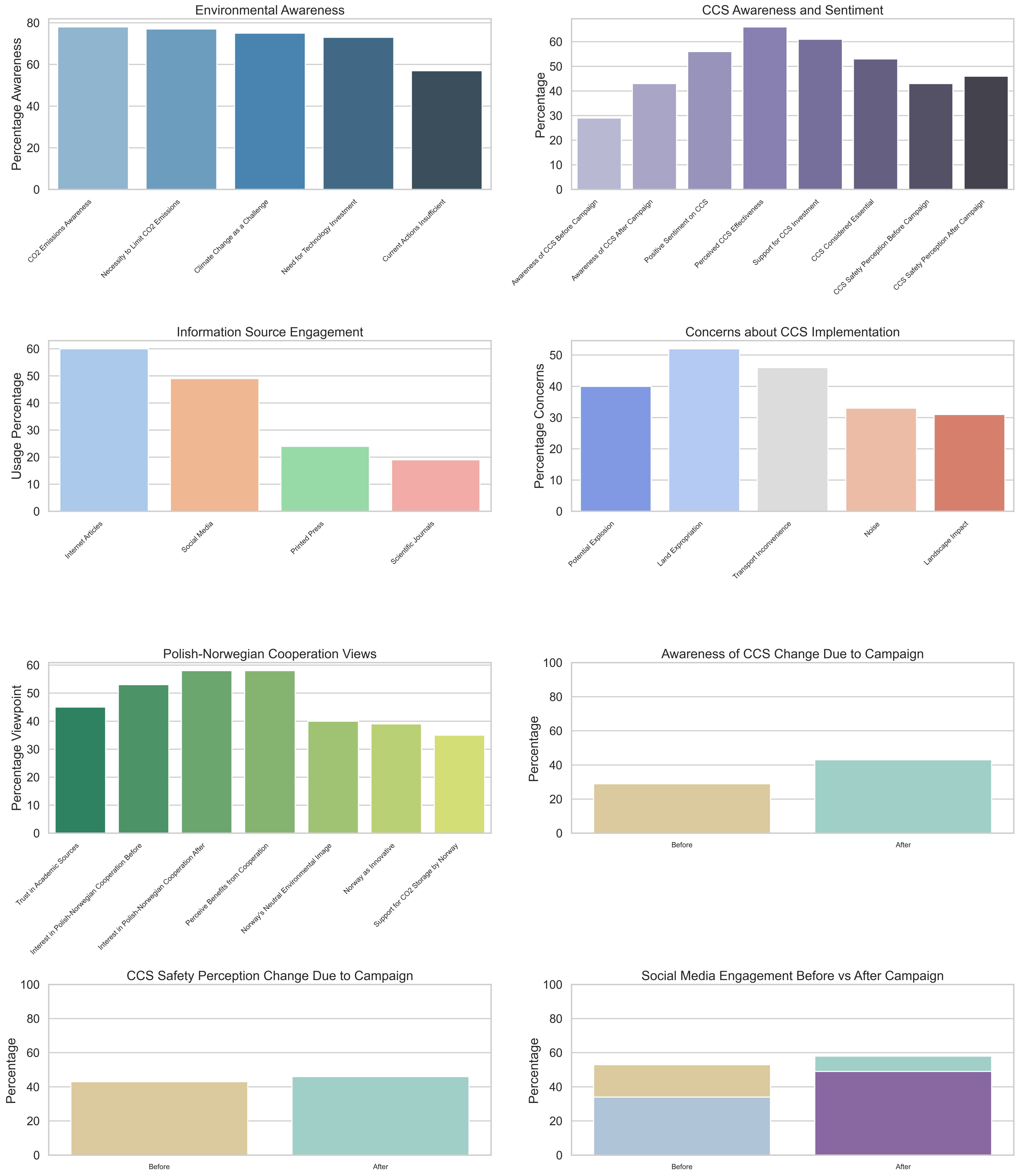}
    \caption{Distribution and insights from pre- and post-campaign surveys, displaying multiple facets of Polish public perception and attitudes regarding environmental awareness and CCS technology. 
        Subplots show: 
        (a) public awareness levels about CO$_2$ emissions and technology investments required, 
        (b) sentiments and safety perceptions about CCS before and after an informational campaign, 
        (c) engagement with various information sources,
        (d) concerns regarding the practical implementation of CCS projects, 
        (e) perceptions and views on international cooperation related to CCS, particularly with Poland and Norway, 
        and (f-h) changes in awareness and sentiment as well as source engagement before and after information campaigns. 
        The results highlight significant perceptual shifts post-campaign and indicate major areas of public concern and interest.}
    \label{fig:figure6}
\end{figure}

\subsection{Awareness, attitudes, and perception}
The surveys highlight that a significant portion of Polish respondents are aware of the detrimental effects of carbon dioxide on atmospheric conditions. As Figure \ref{fig:figure6} demonstrates, a substantial 78\% acknowledge the critical role of CO$_2$ emissions in climate change, with 77\% supporting the reduction of emissions, including CO$_2$ and other greenhouse gases, to preserve environmental health. Climate change poses a considerable challenge, recognized by 75\% of respondents, and there is a perceived need for investment in CO$_2$ capture technologies, with 73\% expressing this viewpoint. Despite recognizing governmental endeavors in emission reduction, 57\% find these efforts inadequate and advocate for new solutions, although only 40\% show awareness of CCS methods as a viable remedy (Fig. \ref{fig:figure6}).

Interest in climate change and ecological issues is expressed by 68\% of the populace, with 62\% sourcing information online. This behavior indicates a broad understanding of CO$_2$'s negative impacts and the necessity for proactive measures. Among those interested, 15\% engaged with related content multiple times weekly in the past six months, with a 46\% frequency of weekly encounters. Online articles (60\%) and social media (49\%) are principal sources of information, overshadowing printed press (24\%) and scientific publications (19\%). Despite 40\% being familiar with CCS potential, just 30\% recognize the term CCS. Additionally, 47\% remain unsure of the specific sectoral applications of this technology. Nevertheless, 66\% trust in CCS's efficacy against climate change, and 53\% view it as imperative to meet the EU's emission targets. Yet, opinions diverge on the suitability of geological conditions (44\%) and technology safety (43\%), signaling the need for public education to address societal apprehensions.

\subsection{Implementation challenges and outlook}
CCS garners support for installation near industrial zones, but its application near residential areas reveals mixed sentiments. Industrial CO$_2$ capture facilities receive a favorable reception, yet underground storage invites varied responses, with 28\% approval and 32\% opposition. Concerns are fuelled by insufficient information and fears of unspecified risks, including possible land expropriation (52\%), increased heavy transport traffic (46\%), and CO$_2$ explosion threats (40\%), underlining educational gaps. Local impacts such as noise (33\%) and visual landscape changes (31\%) are seen as less problematic (Fig. \ref{fig:figure6}).

An analysis of trends between survey phases shows increased recognition of CCS technologies, influenced by pre- and post-campaign comparisons. Awareness rose from 29\% to 43\%; two-thirds of surveyed individuals encountered CCS information in media or public spaces, often via social media (34\%), television (34\%), and online news (33\%). However, clear association with economic sectors poses a challenge, as 60\% are uncertain of CCS applications, up from 47\%. Positivity toward CCS persists among recent information consumers, with 56\% displaying favorable attitudes and 66\% advocating its investment. Recognition of CCS safety grew from 43\% to 46\%, and 55\% expressed interest in expanding their knowledge.

Scientific journals are hailed as the most reliable CCS information source by 45\% of Polish respondents, surpassing television (24\%), trade media (24\%), and online news (21\%). The academic community is trusted significantly more (76\%) than industry (35\%), NGOs (19\%), or government (18\%), suggesting that CCS information proliferation should primarily occur through scientific channels. 

Furthermore, as shown in Figure \ref{fig:figure6}, Polish-Norwegian cooperation in CCS is favorably regarded, with 58\% perceiving more advantages than drawbacks, a rise from 53\%, and reduced indecision to 20\% from 26\%. Awareness of mutual benefits, such as Norway receiving Poland's CO$_2$, increased to 39\% from 35\%. The reduced perception of asymmetric benefits for Poland and the increased belief in the balanced advantages of Norwegian decarbonization involvement highlights a promising collaborative landscape.

\section{Discussion on Pathways to Strategic Partnerships}

\subsection{R\&D and Academic Training}

Joint Research and Development (R\&D) activities have emerged as critical components for the success and sustainability of the Norwegian-Polish CCS partnership. These collaborations focus strategically on site-specific considerations essential for facilitating CO$_2$ storage pilots and full-scale sequestration within Poland's geological formations. Innovation in carbon capture techniques, the advancement of existing technologies, and cooperative testing or piloting at facilities like the Technology Center Mongstad (refer to (Fig. \ref{fig:figuremap}) in Norway are proposed to align with local industry requirements.

Stakeholder feedback and analysis of different activities have further revealed the importance of involving the academic and R\&D sectors across the entire collaborative spectrum. Their central role can ensure continuity, fosters open collaboration, and promotes an organized and systematic approach to sustaining and further developing bilateral contacts and planning. Under these assumptions, the University of Oslo, renowned for leading CCS-related activities within Environmental Geosciences, and AGH University of Krakow, with its Energy Center, were strategically positioned to initiate and advance this initiative. These two closely collaborating parties are focused on facilitating future phases with increased involvement from the industry, public, and policy sectors.

The partnership's commitment to academic exchanges, joint Ph.D. programs, and specialized training courses underscores its potential to substantially enhance academic and professional workforce capabilities. Moreover, establishing initiatives similar to Norway's Centres for Environment-friendly Energy Research (FME) is envisaged as a key Norwegian-Polish collaborative effort. These measures aim to nurture a skilled workforce proficient in managing the comprehensive CCS value chain, thereby ensuring long-term sustainability and operational efficiency.

The academic and R\&D sectors are integral to any pilot or industrial studies, particularly where technological or field-related questions arise. These questions necessitate collaboration with competent researchers to delve into and support industrial endeavors. Moreover, these sectors can significantly contribute to studies related to hub formation, technology retrofitting, geological-site subsurface evaluations, transport planning, carbon accounting, and more, thereby enhancing the overall efficacy and breadth of CCS initiatives.

\subsection{Industrial Synergies and Joint Ventures}

Through this collaboration, opportunities for industrial synergies and technology transfer between Norwegian and Polish entities abound. Sharing expertise in project operations, maintenance, and joint development enhances Polish industries' capabilities, enabling them to leverage Norwegian insights into CCS technology while mitigating associated risks. Establishing joint ventures and partnerships accelerates project deployment and opens new business avenues, aligning both countries with the rising demand for CCS facilities.

Besides joint industrial collaboration for developing capture hubs and providing fit-for-purpose capture technology services and consultations to emitters, an essential aspect is facilitating swift internalization of emission reduction requirements. CO$_2$ storage collaborations emerge as a critical focal area. Until joint efforts, along with contributions from the R\&D sector as mentioned in an earlier section, yield results through necessary pre-assessments, site-specific characterization, and pilots in Poland saline aquifers and hydrocarbon fields. The available storage pore volume in the Norwegian Continental Shelf (NCS), under recently awarded North Sea storage licenses (Fig. \ref{fig:figuremap}), can serve as an immediate and viable option for transporting and storing Polish carbon emissions.

Moreover, developments in the Gdansk area aimed at establishing a terminal (Fig.\ref{fig:figure2}) can align seamlessly with the Long Ship and similar initiatives to jumpstart CCS deployment. This coordination ensures that captured CO$_2$ can be safely and readily stored, enhancing the feasibility and attractiveness of CCS projects.

A cornerstone of the initiative is the inclusive and comprehensive development of infrastructure necessary for the effective transport and storage of CO$_2$ across Poland. Conducting detailed feasibility assessments of various CO$_2$ transport modalities—such as pipelines, railways, inland waterways, and export ports for international shipping facilities is paramount. The potential for repurposing existing pipelines or creating new routes along current corridors offers opportunities to optimize infrastructure usage.

These detailed feasibility studies, conducted in collaboration with the research and development (R\&D) sector, assess technical viability while exploring a range of scenarios. This integration of diverse solutions forms a key component of the national CCS strategy, ensuring that infrastructure development is both strategic and forward-looking, thereby supporting Poland's climate mitigation goals.

\subsection{Public engagement and stakeholder participation}

Active public engagement and stakeholder participation are indispensable for the successful deployment of CCS technologies (Fig. \ref{fig:figure4})). This engagement is not only about disseminating information but also about creating a continuous dialogue that addresses public concerns and builds trust. Drawing from current campaign experiences (discussed above) and stakeholder workshops, it is evident that education and awareness initiatives serve as fundamental tools in combating misinformation, which can otherwise hinder the acceptance and implementation of CCS projects.

A well-structured public campaign should leverage various communication channels to reach diverse demographic groups, emphasizing the importance of CCS in mitigating climate change and meeting emission reduction targets. This is particularly effective when initiated early, incorporating educational programs within school curriculums to instill an understanding of environmental challenges among younger generations. Such early engagement ensures a culture of awareness and responsibility, which is crucial for long-term support.

Public campaigns should also focus on explaining the technological processes, safety measures, and potential economic benefits of CCS. Demystifying the technology and showcasing successful examples, such as those in Norway and other regions, can boost public confidence. The use of relatable and accessible language in these campaigns will help bridge the gap between scientific complexities and public understanding.

Moreover, sustainable public support is fostered through transparency and the inclusion of public feedback in decision-making processes. Facilitating open dialogues with local communities and stakeholders not only resolves potential issues but also cultivates a sense of ownership and involvement among the populace. By addressing specific local concerns, such as environmental safety and economic impacts, these dialogues can remove barriers to acceptance.

Poland can adopt similar strategies based on Norway’s successful public communication practices, which have demonstrated efficacy in building stakeholder confidence. Norwegian experiences highlight the importance of harnessing trusted sources of information, such as scientific journals and academic channels, to disseminate accurate and reliable CCS data. Such strategic communication efforts ensure that public engagement is not a one-time initiative but an ongoing process that evolves with the project’s development.

\subsection{Legal, regulatory, and economic frameworks}
Harmonizing legal and regulatory frameworks is essential for achieving seamless cross-border operations. Key to this endeavor is the simplification of permitting processes and the alignment of national strategies with overarching EU climate objectives. The integration of CCS initiatives into the EU Emissions Trading System (ETS) further streamlines these efforts. Establishing clear liability frameworks and minimizing bureaucratic obstacles are critical to creating a conducive policy environment. In addition, implementing financial incentives—such as carbon contracts for difference (CCfDs), investment grants, and tax benefits from both Polish and Norwegian governments—significantly enhances the economic viability and market accessibility of CCS technologies. These combined efforts facilitate a robust and efficient deployment of CCS, ultimately contributing to the broader goal of sustainable climate action across Europe.

While the Norwegian-Polish CCS network holds immense promise, it is not without its challenges. These include geopolitical risks, potential project delays, and ever-evolving regulatory landscapes, all of which necessitate vigilant and proactive management. To adeptly navigate these complexities, it is imperative to prioritize continuous monitoring and the implementation of adaptive strategies. By integrating comprehensive risk assessments and mitigation plans into the overall strategic framework, the network can ensure resilience and maintain flexibility, thereby facilitating robust execution.

This collaborative endeavor signifies a pivotal advancement towards achieving the European Union's climate objectives, positioning both Norway and Poland as influential players in shaping the global CCS agenda. The partnership not only catalyzes economic development but also expedites the transition to a low-carbon future. Embracing these strategic imperatives enables the Norwegian-Polish CCS network to effectively navigate uncertainties while maximizing its impact on sustainable climate policy and innovation.

\section*{Concluding Remarks}

The European Green Deal outlines the pursuit of climate neutrality by 2050, which makes Carbon Capture and Storage (CCS) crucial for decarbonizing hard-to-abate sectors. The Norwegian-Polish CCS collaboration exemplifies the transformative potential of international partnerships in tackling decarbonization challenges, leveraging each nation's unique strengths.

Norway's pioneering role in CCS is showcased through extensive projects providing insights into the technical and economic facets of CO$_2$ sequestration. The Northern Lights initiative further establishes Norway as a global leader, supported by strong legislative frameworks and investments in research and development.

While Poland is in the early stages of CCS deployment, it presents significant potential for growth. Recent legislative changes foster an environment ripe for CCS development, aiding feasibility studies that leverage Poland's geological storage capacities. However, challenges persist in customizing solutions for industrial emissions and creating dedicated CO$_2$ transport infrastructure, necessitating robust regulatory frameworks for permitting, liability management, and public engagement.

This network serves as a model for international collaboration, demonstrating how synergistic expertise and coordinated efforts can expedite global CCS deployment. Knowledge sharing and industrial synergies facilitate technology transfer and operational improvements, while public engagement and education address concerns and build trust. Harmonizing legal and economic frameworks is essential for seamless cross-border operations. Integrating strategies with EU climate objectives, including financial incentives, enhances CCS viability. 

The academic and R\&D sectors can play an integral role in ensuring a systematic approach to collaboration, addressing technological challenges, and evaluating site and transport plans. The University of Oslo and AGH University of Krakow lead these efforts, fostering innovation and continuous learning.

The success of this network hinges on sustained political commitment, strategic infrastructure investments, and public trust. This partnership marks significant progress toward the European Union’s climate goals, positioning Norway and Poland as key players in the global CCS agenda. Ultimately, integrating CCS into broader climate strategies is critical for effective climate action, paving the way for a sustainable, low-carbon future.

\section*{Acknowledgments}
The authors acknowledge the support and funding received from the "Norwegian-Polish CCS Network: Acceleration of Climate Change Mitigation Technologies Deployment" project, funded through the EEA and Norway Grants under the Green Transition Collaborations Programme (FWD-Green-11). This research was facilitated by the collaborative efforts of the AGH University of Krakow, University of Oslo, and Norwegian Energy Partners (NORWEP).

\subsection*{CRediT authorship contribution} 
Mohammad Nooraiepour: Conceptualization, manuscript design, investigation, data curation and visualization, and writing the original draft. Pawel Gladysz: Writing – review \& editing of final draft. Eirik Melaaen: Writing – review \& editing of final draft. 

\subsection*{Conflicts of Interest}
The authors declare no conflict of interest regarding the publication of this article.



\printbibliography

\end{document}